\documentstyle{article}

\title{\bf Evolution of the gluon density in $x$ with a
running coupling constant}
\author{Mikhail Braun \thanks{
 Permanent address: Dep. High-Energy Physics, University of St. Petersburg,
198904 St.Petersburg, Russia}
, Gian Paolo Vacca  \\
Department of Physics, University of Bologna\\
Istituto Nazionale di Fisica Nucleare - Sezione di Bologna.}
\def\beq{\begin{equation}}
\def\eeq{\end{equation}}
\def\bea{\begin{eqnarray}}
\def\eea{\end{eqnarray}}
\def\noi{\noindent}
\def\kt{k_{\bot}}

\begin{document}
\maketitle
\medskip
\centerline{\bf Abstract.}
This is a draft describing the calculation of the evolution of the 
gluon density in $x$ from an initial value $x=x_{0}=0.01$ to smaller
values, up to $x=10^{-8}$ in the hard pomeron formalism with a running
coupling introduced on the basis of the bootstrap equation.
The obtained gluon density is used to calculate the singlet part of the
proton structure function. Comparison with experiment and the results
following from the fixed coupling evolution is made. 

\newpage
\section{ Introduction.}

Recent results obtained at HERA \cite{HERA,ZEUS2} may be interpreted as a 
manifestation
of the hard pomeron, which naturally explains a sharp rise of $F_{2}(x,Q^{2})$
at low $x$. The original BFKL hard pomeron, however, has a drawback of
treating the coupling constant as fixed, since it sums only powers of 
$\log 1/x$ and not those of $\log Q^2$. A rigorous way to introduce a 
running coupling into it still remains beyond the possibilities of the 
theory, since it inevitably involves a problem of low $Q^2$ behaviour 
and thus of confinement. In a series of papers \cite{braun1,braun2,bvv,bv2}
we have adopted a 
more intuitive way to attack this problem, based on the so-called 
bootstrap relation \cite{lip1,bart1}, which is, in fact, the unitarity 
condition for 
the $t$-channel with a colour quantum number of a gluon. It ensures that
the one-reggeized-gluon exchange supposed to give a dominant 
contribution in this channel is unitary by itself, which is a necessary 
requirement to use it as an input for the construction of the BFKL 
pomeron.

Assuming that this fundamental requirement should be preseved in the 
theory with a running coupling, we proposed a minimal modification of the 
BFKL pomeron equation which, on the one hand, satisfies the bootstrap 
condition and, on the other hand, leads to the standard results with the 
running coupling in the double log (DL) limit, i.e. when leading terms 
in the product $\log1/x\log Q^2$ are summed. This modification reduces 
to  the substitution of every momentum squared $k^2$ in the pomeron
equation by a function $\eta(k)$ which at large $k$ behaves as
$k^2/2\alpha_{s}(k^{2})$. The behaviour of $\eta(k)$ at low $k$ remains 
beyond any theoretical controle. We parametrize it as
\beq
\eta(k)=(b_{0}/8\pi)(k^{2}+m^{2})\ln\frac{k^{2}+m^{2}}{\Lambda^{2}}
\eeq
where $\Lambda$ is the standard QCD parameter,
$b_{0}=11-(2/3)N_{f}$ and the effective gluon 
mass $m$ simulates both the confinement effects and the freezing of the 
coupling.

Solving numerically the pomeron equation in this approach we found two 
supercritical pomerons \cite{bvv}. Adjusting the mass $m$ to fit the 
experimental slope of the leading pomeron of 0.25 $(GeV/c)^{-2}$ we 
obtained for their intercepts
\[\Delta_{0}=0.384,\ \ \ \Delta_{1}=0.191\]
and the slope of the subdominant pomeron results $\alpha'_{1}=0.124\ 
(GeV/c)^-2$. Calculating observable quantities with only these two
asymptotic states taken into account we found that the picture which 
emerges, in all probability, corresponds to energies much higher than 
the present ones \cite{bvv}. In particular the average $\langle\kt\rangle$ was 
found to be very large ($\sim 10\ GeV/c$) and independent of energy, 
which may indicate a saturation of its growth observed at present 
energies \cite{bv2}.

To describe the present experimental data it is then necessary to take 
into account all the states from the pomeron equation spectrum. This can 
be achieved by converting the pomeron equation into an evolution 
equation in $1/x$ and solving it with an initial condition at some
(presumably small) value $x=x_{0}$. In such an approach, taking a 
nonperturbative input at $x=x_{0}$ adjusted to the experimental data,
also the problem of coupling the pomeron to the hadronic target is 
solved in an effective way.

This note is devoted to realizing such a program. In Sec. 2 we state our 
basic equations. The most difficult part of the program is to pass from 
the gluon density to the observable structure function. It is discussed 
in Sec. 3. Sec. 4 is devoted to fixing the initial gluon distribution
for the future evolution. In Sec. 4 we present our numerical results. 
Sec. 5 contains a discussion and some conclusions.
\section{Basic equations}
For the forward scattering amplitude the pomeron equation reads
\beq
(H-E)\psi=\psi_{0}
\eeq
Here $\psi$ is a semi-amputated (one leg amputated only) pomeron wave 
function; $E=1-j$ is the pomeron "energy", related to its complex 
angular momentun $j$; $H=T+V$ is the "Hamiltonian" consisting of the 
kinetic energy given by the sum of the two gluon Regge trajectories,
$T=-2\omega$ and of the potential energy $V$. With a running coupling 
introduced according to \cite{braun1,braun2} both are expressed via 
the mentioned function  $\eta$ (Eq. (1));
\beq
T(k)=\frac{N_c}{(2\pi)^2} \int\frac{d^2k'\eta(k)}{\eta(k')\eta(k-k')}
\eeq
and
\beq
V\psi(k)=-\frac{T_1 T_2}{(2\pi)^2} \int\, d^2k'\psi(k')\left(\frac{2}
{\eta(k-k')}- \frac{\eta(0)}{\eta(k)\eta(k')}\right)
\eeq
where $N_c$ is the number of colours and $T_{1(2)}$ are the colour
operators for the two interacting gluons; in the vacuum channel we
have $T_1T_2=-N_c$.
Finally, the inhomogeneous term $\psi_{0}$ represents the interaction 
vertex  between the pomeron and the hadronic target.

Taking the Mellin transformation of (2) one converts  it into an 
evolution equation in $1/x$:
\beq
\frac{\partial}{\partial\ln 1/x}\psi(x,k)=-H\psi(x,k)
\eeq
which should be supplemented with an initial condition at some $x=x_{0}$
\beq
\psi(x_{0},k)=\psi_{0}(k)
\eeq
containing the nonperturbative input about the coupling to the hadronic 
target.

The physical interpretation of the pomeron wave function is provided by 
the fact that in the DL approximation Eq. (5) reduces to an equation for 
the fully amputated function $\phi(x,k)=\eta(k)\psi(x,k)$:
\beq
\frac{\partial}{\partial\ln k^2}
\frac{\partial}{\partial\ln 1/x}\phi(x,k)=\frac{3\alpha_{s}(k^{2})}
{\pi}\phi(x,k)
\eeq
which coincides with the standard equation for the unintegrated 
gluon density $xg(x,k^{2})$ in the DL limit. In fact, this circumstance 
lies at the root of our method to introduce a running coupling into the 
scheme. Thus we may identify
\beq
\phi(x,k)=cxg(x,k)
\eeq
The normalizing factor $c$ cannot be determined from the asymptotic 
equation (7). We shall be able to fix it by studying the coupling of the 
pomeron to the incoming virtual photon in the next section.
\section{Coupling to the virtual photon}
Once the function $\phi$ proportional to the gluon density is 
determined, one has to couple it to the projectile particle to calculate 
observable quantities. In particular, to find the structure function of 
the target one has to couple the gluons to the incoming virtual photon, 
that is, to find the colour density $\rho(q,k)$ which connects the photon 
of momentum $q$ to the gluon of momentum $k$. This problem is trivial 
within the BFKL approach with a fixed small coupling. Then it is 
sufficient to take the colour density in the lowest order, which 
corresponds to taking for it the contribution of a pure quark loop
into which the incoming photon goes
$\rho_{0}(q,k)$.

The problem complicates enormously when one tries to introduce a running 
coupling into $\rho$. Then one has to take into account all additional 
gluon and $q\bar q$ pair emissions which supply powers of the logarithms
of transverse momenta. Apart from making the coupling run, they will 
evidently change the form of $\rho(q,k)$. Unfortunately the bootstrap 
relation can tell us nothing about the ultimate form of the colour 
density with a running coupling, which essentially belongs to the 
$t$-channel with a vacuum colour quantum number. So we have to find a 
different way to introduce a running coupling into $\rho$.

A possible systematic way to do this consists in
applying to the photon-gluon coupling the DGLAP evolution equation.
One may separate the colour 
density from the rest of the amplitude by restricting its rapidity range 
to some maximal rapidity $y_{0}\sim\log Q^2$ (which, of course should be
much smaller than the overall rapidity $Y\sim\log Q^{2}/x$). Then the 
kinematical region of $\rho(q,k)$ will admit the standard DGLAP 
evolution in $Q^2$. Solving this equation one will find the quark 
density at scale $Q^2$ of the gluon with momentum $k$ (i.e. essentially 
the structure function of the gluon with the virtuality $k^2$). This is 
exactly the quantity needed to transform the calculated gluon density 
created by the target into the observable structure function of the target.
As a starting point for the evolution one may take the perturbative 
colour density $\rho_{0}$ at some low $Q^2$ when the logs of the 
transverse momenta might be thought to be unimportant.

This ambitious program, combining both evolution in both $1/x$ and 
$Q^2$, does not, however, look very simple to realize. As a first step, 
to clearly see the effects of the introduction of a running coupling 
according to [ 3,4], we adopt a more phenomenological approach here, 
trying to guess a possible correct form for $\rho(q,k)$ on the basis of 
simple physical reasoning and also using the DL approximation to fix
its final form.

With a pure perturbative photon colour density one would obtain for the 
$\gamma^{*}p$ cross-section
\beq
\sigma(x,Q^{2})=\int\frac{d^{2}k\rho_{0}(q,k)\phi(x,k)}{(2\pi)^{2}\eta^{2}(k)}
\eeq
In fact, the projectile particle should be coupled to the full pomeronic 
wave function $\phi/\eta^2$. From the physical point of view this 
expression is fully satisfactory for physical particles. However it is 
not for a highly virtual projectile.

To see this, we first note that for the forward amplitude our method of 
introducing a running coupling reduces to a very simple rule: the scale 
at which the coupling should be taken is given by the momentum of the 
emitted real gluon ($(k-k')^2$ in the upper rung in Fig. 1). Now take 
$Q^{2}$ very large and apply the DL approximation. Then the momenta in 
the ladder  become ordered from top to bottom
\[Q^{2}>>k^{2}>>{k'}^{2}>>.....\]
In this configuration, as can be traced from (2) and (9), all $\alpha_{s}$'s
acquire the right scale (i.e. corresponding to the DGLAP equation) except
for the upper rung: $\alpha_{s}(k^{2})$ appears twice. This defect can be 
understood if one notices that the upper gluon is, in fact, coupled to a 
virtual particle. If this particle were a gluon, then the interaction (4) 
would cancel one of the two $\alpha(k^{2})$'s and substitute it by an 
$\alpha$ taken at the scale corresponding to its own virtuality.
We assume that something similar should take place also for virtual 
quarks to which the gluon chain may couple. The scale of the particle
momenta squared
which enter the upper blob in Fig. 1 should have the order $Q^{2}$ (this
is the only scale that remains after these momenta are integrated out).
As a result  the lowest order density should be rescaled 
according to
\beq
\rho_{0}(q,k)\rightarrow\frac{\alpha(Q_{1}^{2})}{\alpha(k^{2})}\rho_{0}(q,k)
\eeq
where $Q_{1}^{2}$ has the same order as $Q^2$.

The approximation we assume in this paper is that the substitution 
(10) is sufficient to correctly represent the photon colour density with a 
running coupling. We shall check its validity by studying the quark density 
which results from (10) in the DL approximation
and comparing it with the known result 
based on the DGLAP equation.

Explicitly the zeroth order density $\rho_{0}$ has the following forms 
for the transverse (T) and longitudinal (L) photons (see e.g. \cite{nz1} and do 
the integration in the quark loop momenta)
\beq
\rho^{(T)}_{0}(q,k)=\frac{3e^{2}}{8\pi^2}\sum_{f}Z_{f}^{2}
\int_{0}^{1}d\alpha\left((\alpha^{2}+(1-\alpha)^{2})
((1+2z^{2})g(z)-1)+\frac{\zeta}{\alpha(1-\alpha)+\zeta}(1-g(z))\right)
\eeq
\beq
\rho^{(L)}_{0}(q,k)=\frac{3e^{2}}{2\pi^2}\sum_{f}Z_{f}^{2}
\int_{0}^{1}d\alpha\frac{(\alpha(1-\alpha))^{2}}{\alpha(1-\alpha)+\zeta}
(1-g(z))
\eeq
Here the summation goes over the quark flavours. The dimensionless 
variables $\zeta$ and $z$  are defined as
\beq
\zeta=\frac{m_{f}^{2}}{Q^2},\ \ 
z=\frac{k^2}{4Q^2}\frac{1}{\alpha(1-\alpha)+\zeta}
\eeq
and $m_{f}$ and $Z_{f}$ are the mass and charge of the quark of 
flavour $f$. The function $g(z)$ is given by
\beq
g(z)=\frac{1}{2z\sqrt{z^{2}+1}}\ln\frac{\sqrt{z^{2}+1}+z}{\sqrt{z^{2}+1}-z}
\eeq
 The structure function is obtained from the cross-section 
by the standard relation
\beq
F_{2}(x,Q^{2})=\frac{Q^2}{\pi e^2}(\sigma^{(T)}+\sigma^{(L)})
\eeq

In the DL limit only the transverse cross-section contributes. We can 
also neglect the quark masses in this approximation. Then, with a 
substitution (10), from (9), (11)  and (15) 
we obtain an expression for the quark (sea) density of the target
\beq
xq(x)=\frac{3}{\pi^{2}b_{0}^{2}}\frac{Q^2}{\ln Q^2}
\int^{Q^2}\frac{dk^2\phi(x,k)}{k^{4}\ln k^2}
\int_{0}^{1}d\alpha(\alpha^{2}+(1-\alpha)^{2})
((1+2z^{2})g(z)-1)
\eeq
where $g(z)$ is given by Eq. (14) and we assumed that large values of 
$k^{2}<Q^2$ contribute in accordance with the DL approximation.
In this approximation the asymptotics of the gluon density $xg(x,k^{2})$ 
and consequently of $\phi(x,k^{2})$ is known:
\beq
\phi(x,k^{2})=cxg(x,k^{2})\simeq c
\exp \sqrt{a\ln\frac{1}{x}\ln\ln k^{2}}
\eeq
where $a=48/b_{0}$. Putting (17) into (16), after simple calculations we
find the asymptotical expression for the quark density
\beq
xq(x,k^{2})\simeq\frac{4c}{\pi^{2}b_{0}^{2}}\sqrt{\frac{\ln\ln k^2}
{a\ln 1/x}}\exp \sqrt{a\ln\frac{1}{x}\ln\ln k^{2}}
\eeq
On the other hand, from the DGLAP equation we find, with the same 
normalization
\beq
xq(x,k^{2})\simeq\frac{4}{3b_{0}^{2}}\sqrt{\frac{\ln\ln k^2}
{a\ln 1/x}}\exp \sqrt{a\ln\frac{1}{x}\ln\ln k^{2}}
\eeq
As we observe the approximation (10) for the colour density of the photon 
projectile leads to the correct relation between the quark and gluon 
densities in the DL limit. This justifies the use of (10), at least for
high $1/x$ and $Q^{2}$. Comparing (18) and (19) we also obtain the 
normalization factor $c$ which relates the pomeron wave function to the 
gluon density
\beq
c=\pi^{2}b_{0}/3
\eeq
\section{The initial distribution}
To start the evolution in  $1/x$ we have to fix the initial gluon 
density at some small value $x=x_{0}$. Evidently, the smaller
is $x_{0}$, the smaller is the region
where we can compare our predictions with the experimental data.
On the other hand, if $x_{0}$ is not small enough, application of the
asymptotic hard pomeron theory becomes questionable. Guided by these
considerations we choose $x_{0}=0.01$ as our basic initial $x$
although we also tried $x=0.001$ to see the influence of possible 
subasymptotic effects.

The initial wave function $\phi(x_{0},k^{2})$ has to be chosen in 
accordance with the existing data at $x=x_{0}$ and all $k^2$ available.
The experimental $F_{2}$ is a sum of the singlet and nonsinglet parts, 
the latter giving a relatively small contribution at $x=0.01$. Our 
theory can give predictions only for the singlet part (and one of the 
criteria for its applicability is precisely the relative smallness of 
the nonsinglet contribution). The existing experimental data at $x=0.01$
give values for $F_{2}$ averaged over rather large intervals of $x$ 
and $Q^2$. For all these reasons, rather than to try to adjust our 
initial $\phi(x_{0},k^{2})$ to the pure experimental data, we have 
preferred to match it with the theoretical predictions for the 
gluon density  and the singlet 
part of $F_{2}$ given by some standard parametrization fitted to the 
observed $F_{2}$ in a wide interval of $Q^2$ and small $x$. As such we 
have taken the GRV LO parametrization \cite{grv}. The choice of LO has been 
dictated by its comparative simplicity and the fact that at $x=0.01$ the 
difference betwen LO and NLO is insignificant.

Thus, for the initial distribution we have taken  the GRV
LO gluon density with an appropriate scaling factor. Putting this
density into Eqs. (8),(9) and (15) one should be able to reproduce the 
sea quark density and thus the singlet part of the structure function.
In the GRV scheme the relation between the gluon density and the quark 
density is much more complicated and realized through the DGLAP 
evolution. Since the DGLAP evolution and the pomeron theory are not 
identical, one should not expect that our initial gluon density 
 should exactly coincide with the GRV one to give the same singlet
 structure funcction.  One has also to have 
 in mind the approximate character of our colour density $\rho$ at small
$Q^2$. In fact, with the initial $\phi$ given by (8) and the
gluon density exactly taken from the GRV parametrization at $x=0.01$
we obtain a 30\% smaller values for the singlet part of the structure 
function as given by the same GRV parametrization, the difference 
growing at low $Q^2$. To make the 
description better we used a certain arbitrariness in the scale 
$Q_{1}^{2}$ which enters (10) and also the scale at which the coupling 
freezes in the density $\rho$. The optimal choice to fit the low $Q^2$ 
data is to take
\beq
\alpha (Q_{1}^{2})=\frac{4\pi}{b_0} 
\frac{1}{\ln ((0.17*Q^{2}+0.055\ (GeV/c)^{2})/ \Lambda^{2})}
\eeq
With this $\alpha(Q_{1}^{2})$ the obtained singlet 
structure function at $x=0.01$ 
has practically the same $Q^{2}$ dependence as the GRV one, although it 
results 30\% smaller in magnitude. 
This mismatch can be interpreted in two different ways. Either we may 
believe that the gluon density given by the GRV is the correct one and
the deficiency in the singlet part of the structure function is caused
by our approximate form of the colour density $\rho$ (which is most 
probable). Or we may think that the colour density to be used in the
DGLAP should coincide with ours only for large enough 
$Q^2$ and $1/x$ and at finite values they may somewhat differ (our 
relation (8) was established strictly speaking only in the DL limit).
Correspondingly we may either take the relation (8) as it stands and use 
the GRV LO gluon density at $x=0.01$ in it, or introduce a correcting 
scaling factor 1.3 which brings the structure function calculated with 
the help of (9)-(15) into  agreement with the GRV predictions.
In the following we adopt the second alternative, that is we assume that 
our initial gluon distribution at $x=0.01$ is 30\% higher that the
one given by the GRV parametrization.
The singlet part of the structure function at $x=0.01$
calculated from (9)-(15) with this choice is shown in Fig. 2 together
 with the GRV predictions. 
 However one can easily pass to the 
first alternative by simply reducing our  results by factor 1.3.

\section{Evolution: numerical results}
With the initial wave function $\phi(x=0.01,k^{2})$ chosen as indicated 
in the preceding section we solved the evolution equation for 
$10^{-8}<x<10^{-2}$. 
The adopted calculational scheme was to diagonalize 
the Hamiltonian in (2), reduced to one dimension in the transverse momentum
space after angular averaging, and represent the initial wave function as a 
superposition of its eigenvectors. To discretize $k^2$ a grid was 
introduced, after which the problem is reduced to a standard matrix one.
To check the validity of the obtained results we have also repeated the
evolution using a Runge-Kutta method, resulting in a very good agreement.
The final results obtained for the gluon distribution $xg(x,Q^{2})$ as 
a function of $Q^2$ for various $x$ are shown in Figs. 3 and 4 and as a 
function of $x$ for various $Q^2$  in Figs. 5 and 6. Figs. 3 and 5 
correspond to $x$ and $Q^2$ presently available, whereas Figs. 4 and 6
show the behaviour of the calculated gluon density in the 
region up to very small $x$ and very high $Q^2$, well beyond the present 
possibilities. For comparison we have also shown the gluon densities for the 
GRV LO parametrization \cite{grv}, for the MRS parametrization \cite{mrs} 
and also for the pure BFKL evolution as calculated in \cite{kms}.

Putting the found gluon densities into Eqs. (9)-(15) we obtain the (singlet 
part of) proton structure function $F_{2}(x,Q^{2})$. The results are 
illustrated in Figs. 7 and 8 for the $Q^2$-dependence and Figs. 9 and 10
for the $x$ dependence. As for the gluon densities, the experimentally
investigated region is shown separately in Figs. 7 and 9, in which the
existing experimental data from \cite{ZEUS2} are also presented.

Finally, to see a possible influence of subasymptotic effects, we
have repeated the procedure taking as a starting point for the evolution 
a lower value $x=0.001$. The resulting gluon distributions and structure 
functions are also presented in the above figures.

\section{Discussion and conclusions}
To discuss the obtained results we have to remember that they involve 
two quantities of a different theoretical status. One is the pomeron 
wave function $\phi$ which can be identified with the gluon distribution 
(up to a factor) on a rather solid theoretical basis. The other is
the quark density (which is equivalent to the structure function), for
which we actually have no rule for the introduction of a running 
constant and which in the present calculation involves a 
semi-phenomenological ansatz (10). Evidently the results for the latter
are much less informative as to the effect of the running coupling 
introduced in our way. Therefore we have to separately discuss our
prediction for the gluon distribution, on the one hand, and for
the structure function, on the other.

Let us begin with the gluon distribution. Comparing our results with
those of GRV, which correspond to the standard DGLAP evolution,
we observe that at high enough $Q^2$ and low enough $x$ our distributions
rise with $Q$ and $1/x$ faster than those of GRV. This difference is,
of course, to be expected. The hard pomeron theory in any version
predicts a power rise of the distribution with $1/x$ to be compared with
(19) for the DGLAP evolution. As to the $Q$-dependence, the fixed 
coupling (BFKL) hard pomeron model predicts a linear rise, again
much stronger than (19). Our running coupling model supposedly leads
to a somewhat weaker rise. From our results it follows that it is still
much stronger than for the DGLAP evolution. However one can observe that
these features of our evolution become clearly visible only at quite
high $Q$ and $1/x$. For moderate $Q<10\, GeV/c$ and/or $x>10^{-4}$
the difference between our distributions and those of GRV is 
insignificant. As to the DGLAP evolved MRS parametrization, it
gives the gluon distribution which lies systematically below the
GRV one and, correspondingly, below our values, the difference
growing with $Q$ and $1/x$.

We can also compare our gluon distributions with the pure BFKL evolution
(fixed coupling) results, as presented in [12]. One should note that the 
initial values for the evolution chosen in [12] are rather different
from ours (borrowed from GRV). The initial gluon distribution in [12]
is smaller than ours by a $Q$-dependent factor, equal to $\sim 2.5$
at $Q=2\ GeV/c$ and $\sim 1.4$ at $Q=30\ GeV/c$. If one roughly takes 
that into account then from Fig. 3 one concludes that at moderate $1/x$
our evolution and the pure BFKL one lead to quite similar results.
However at smaller $x$ (Fig. 4) one observes that our running coupling
evolution predicts a weaker rise with $Q$, as expected.

Passing to the structure functions we observe in Fig. 7 and 9 that our 
results give a somewhat too rapid growth with $1/x$ in the region 
$10^{-3}<x<10^{-2}$ as compared to the experimental 
data (and also to the parametrizations GRV fitted to these data). 
With the scaling factor 1.3 introduced to fit the data at 
$x=0.01$ we overshoot the data at $x<10^{-3}$ by $\sim$25\%. Without this 
factor we get a very good agreement for $x<10^{-3}$ but are below 
experiment at $x=0.01$ by the same order. This discrepancy may be 
attributed either to subasymptotic effects or to a poor quality of our 
ansatz (10). Comparison with the result obtained with a lower starting 
point for the evolution $x=0.001$ shows that subasymptotic effects
together with a correct form of coupling to quarks
may be the final answer.

\section {Acknowledgments.}

The authors express their deep gratitude to, Prof G.Venturi for his
constant interest in this work and helpful discussions.
M.A.B. thanks the INFN for its hospitality and financial help during his stay
at Bologna University. 


\section{Figure Captions}

\noi Fig. 1
The forward amplitude for a pomeron coupled to a virtual photon. 

\noi Fig. 2
The singlet part of the structure function of the proton at $x=0.01$. The
continuous line is the result of our calculation while the dashed line
correspond to the GRV prediction.

\noi Fig. 3
 The gluon distributions as a function of $Q^2$ evolved from 
$x=0.01$ and $x=0.001$ for the experimentally accessible kinematical
region. Standard DGLAP evolved parametrizations (GRV-LO and 
MRS) and the BFKL evolved distributions from [12] (we report only few points
connected by lines) are shown for comparison.

\noi Fig. 4
Same as Fig. 3 for asymptotically high values of $Q^2$ and $1/x$.

\noi Fig. 5
 The gluon distributions as a function of $x$ evolved from 
$x=0.01$ and $x=0.001$ for the experimentally accessible kinematical
region. Standard DGLAP evolved parametrizations (GRV-LO and 
MRS) and the BFKL evolved distributions from [12] (we report only few points
connected by lines) are shown for comparison.

\noi Fig. 6
Same as Fig. 5  for asymptotically high values of $Q^2$ and $1/x$.

\noi Fig. 7
$Q^2$ dependence of the singlet part of the proton structure function 
obtained by evolution from
$x=0.01$ and $x=0.001$, compared to the GRV prediction and the ZEUS 94 data.

\noi Fig. 8
Same as Fig. 7 for asymptotically high values of $Q^2$ and $1/x$.

\noi Fig. 9
$x$ dependence of the singlet part of the proton structure function 
obtained by evolution from
$x=0.01$ and $x=0.001$, compared to the GRV prediction and the ZEUS 94 data.

\noi Fig. 10
Same as Fig. 9 for asymptotically high values of $Q^2$ and $1/x$.



\begin{thebibliography}{3}

\bibitem{HERA}{I. Abt et al., H1 Collab.: Nucl. Phys. B407 (1993) 515;\\
	T. Ahmed et al., H1 Collab.: DESY 95-006 (1995);\\
	M. Derrick et al., ZEUS Collab.: Phys. Lett. B316 (1993) 412;\\
	M. Derrick et al., ZEUS Collab.: DESY 94-143 (1994)}
\bibitem{ZEUS2} {ZEUS Collaboration, PREPRINT, Measurement of the $F_2$ 
structure function in deep inelastic 
$e^{+}p$ scattering using 1994 data from ZEUS detector at HERA.}

\bibitem{braun1} {M.A.Braun: Phys. Lett. {\bf B345} (1995) 155.}

\bibitem{braun2} {M.A.Braun: Phys. Lett. {\bf B348} (1995) 190.}

\bibitem{bvv} {M.A.Braun, G.P. Vacca and G. Venturi:
 Phys. Lett. B388 (1996) 823.}

\bibitem{bv2}{M. A. Braun, G. P. Vacca: PREPRINT HEP-PH/9701356,
		accepted for publishing by Phys. Lett. B}

\bibitem{lip1} {L.N.Lipatov: Yad. Fiz. {\bf 23}(1976) 642.}

\bibitem{bart1} {J.Bartels: Nucl. Phys. {\bf B151} (1979) 293.}

\bibitem{nz1}{ N.N.Nikolaev and B.G.Zakharov: Z.Phys. {\bf C49} (1991) 607}

\bibitem{grv}{M. Gl{\"u}ck, E. Reya, A. Vogt: Z. Phys C67 (1995) 433-447}

\bibitem{mrs}{A.D. Martin, W.J. Stirling, R.G. Roberts: Phys. Rew. D50 (1994)
6734}

\bibitem{kms} {J. Kwiecinski,A.D. Martin,P.J. Sutton: Phys. Rew. D44 (1991)
2640.}

\end{thebibliography}
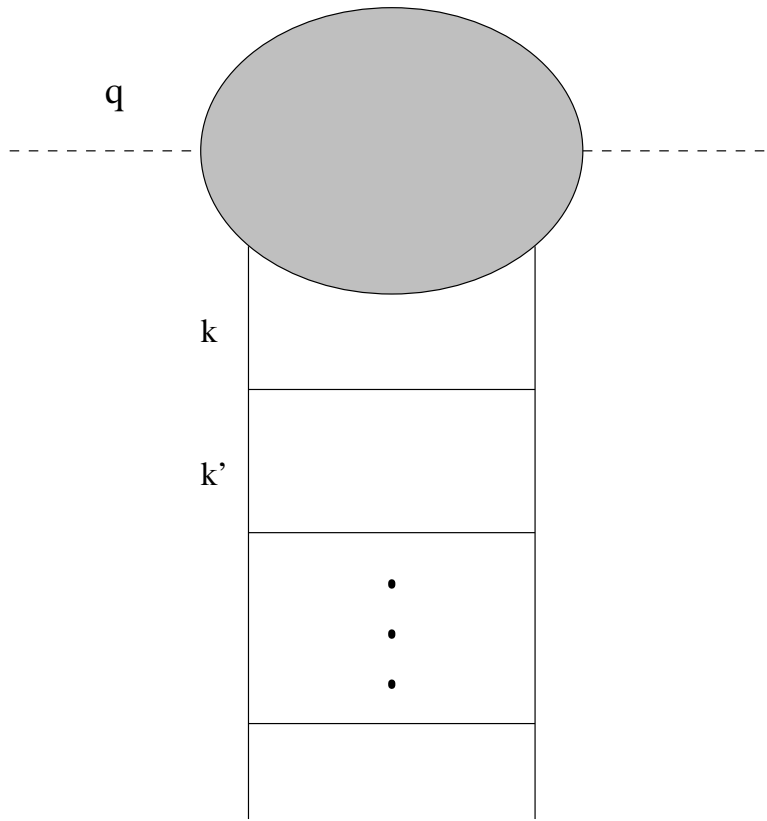
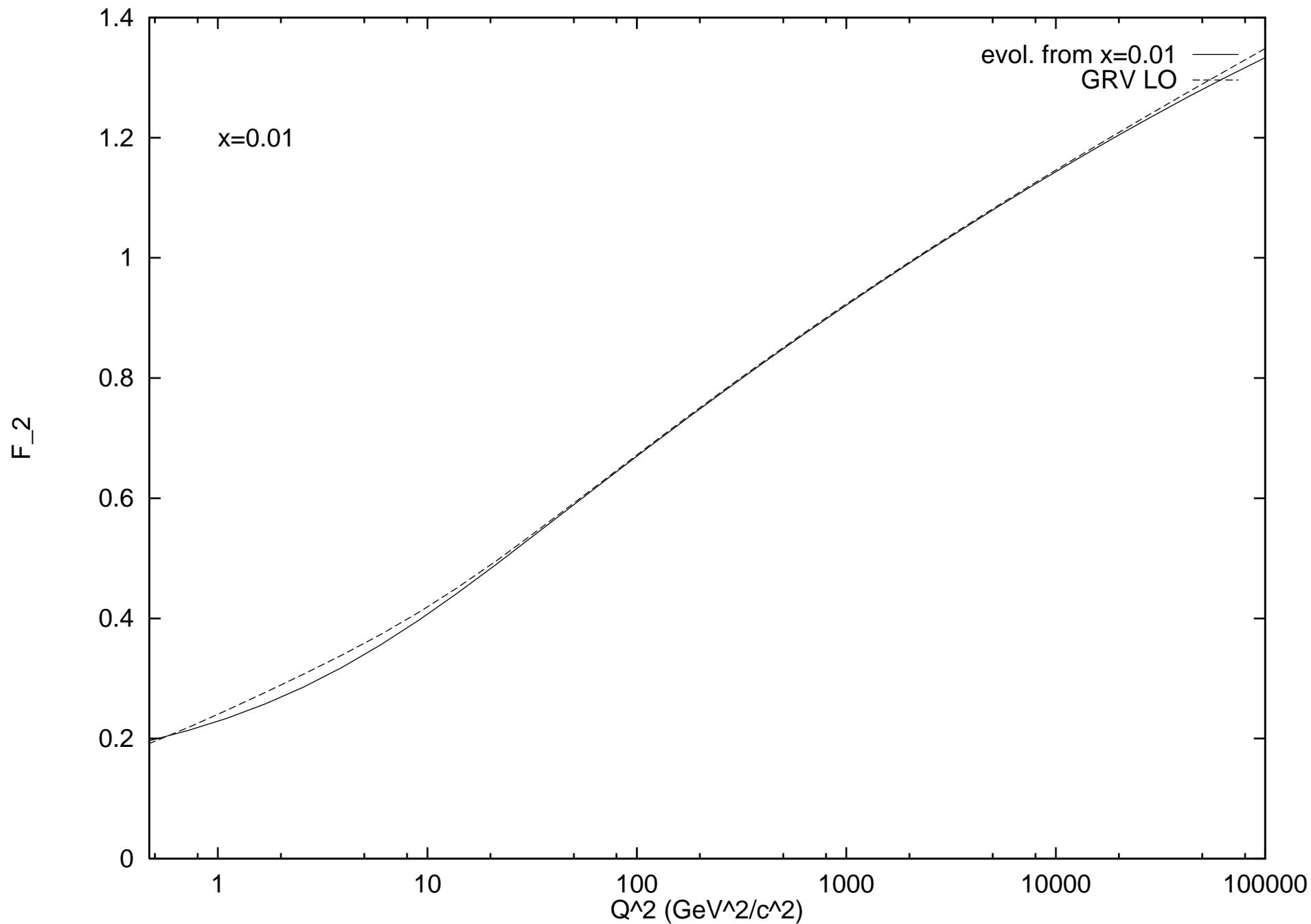
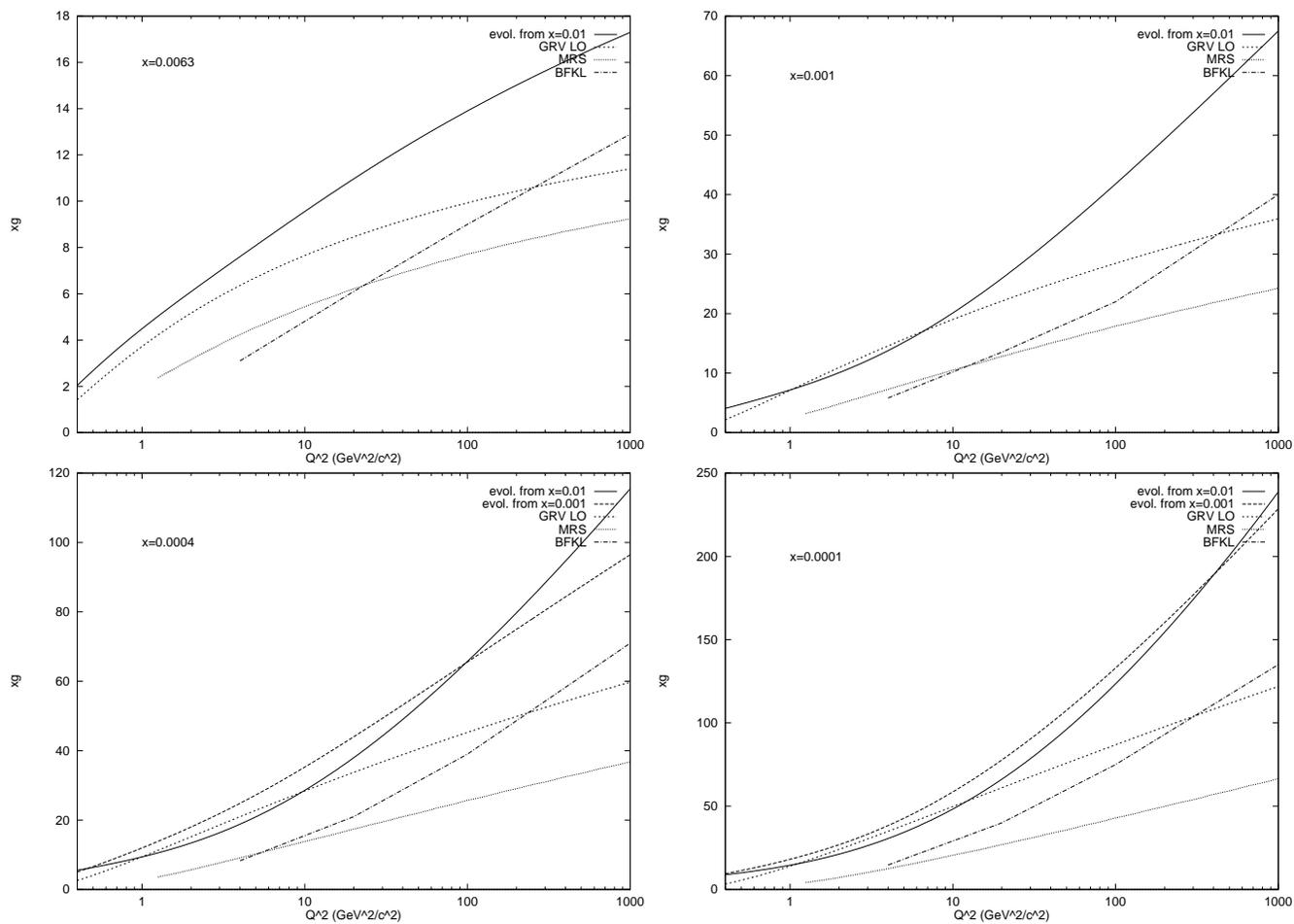
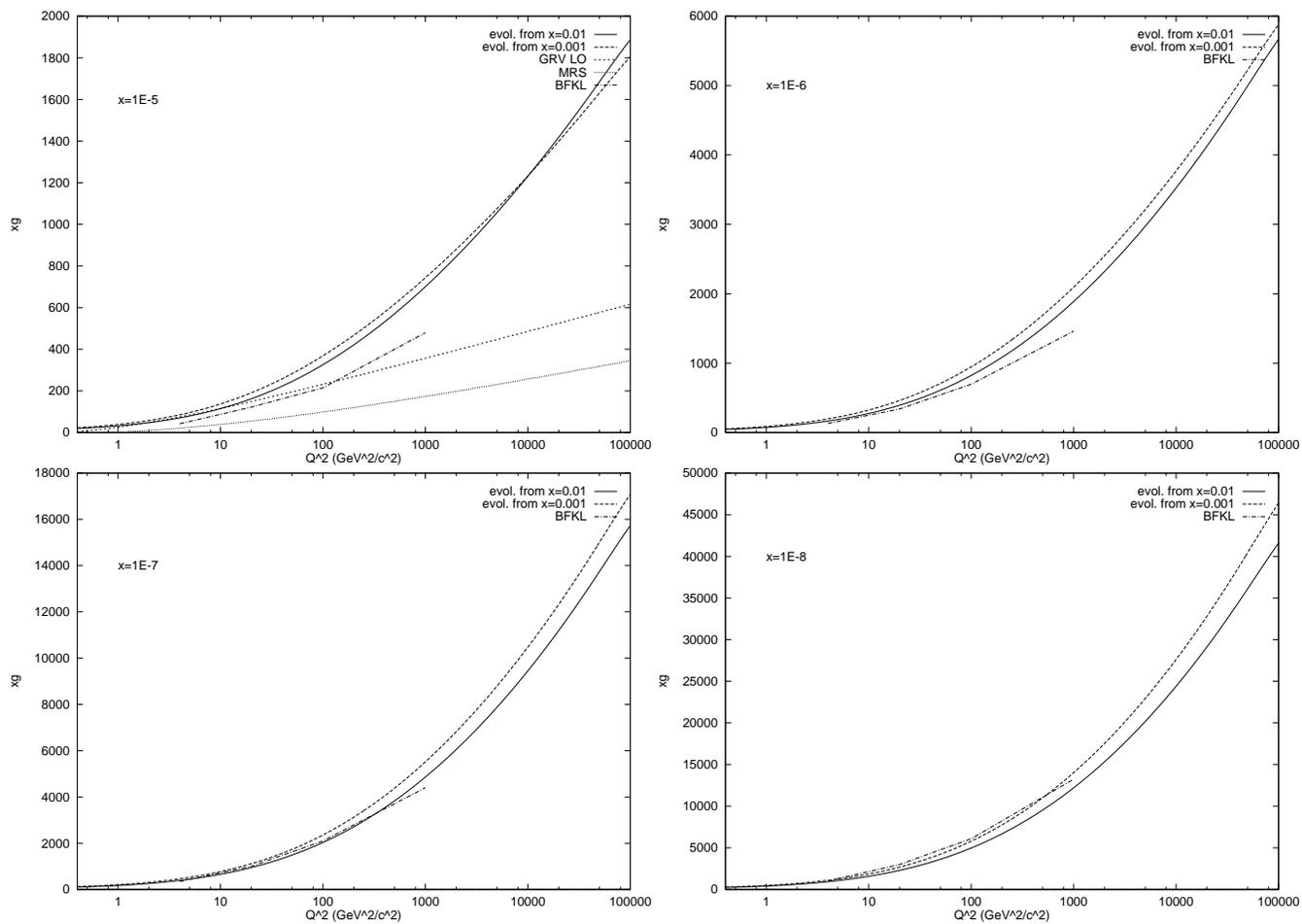
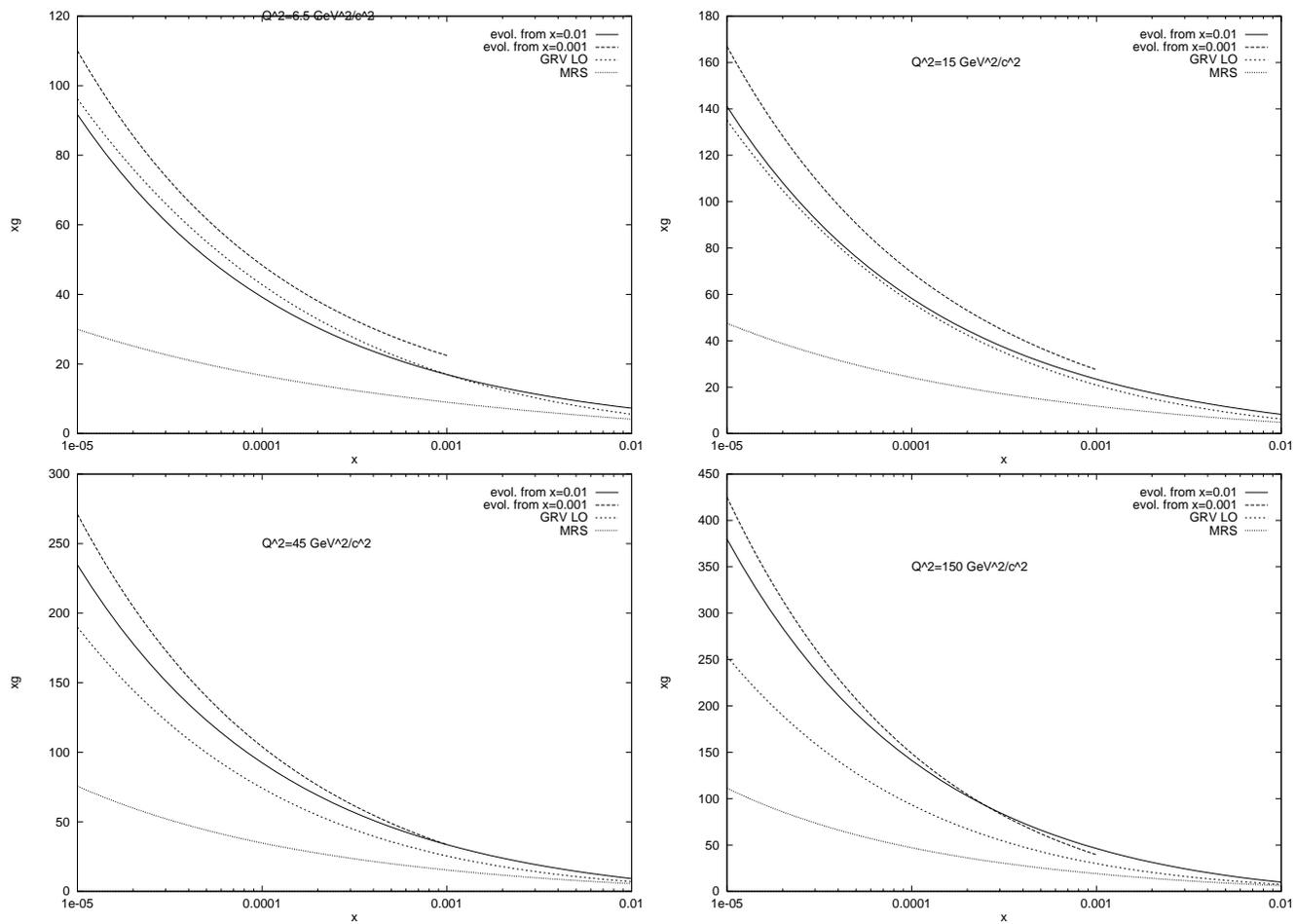
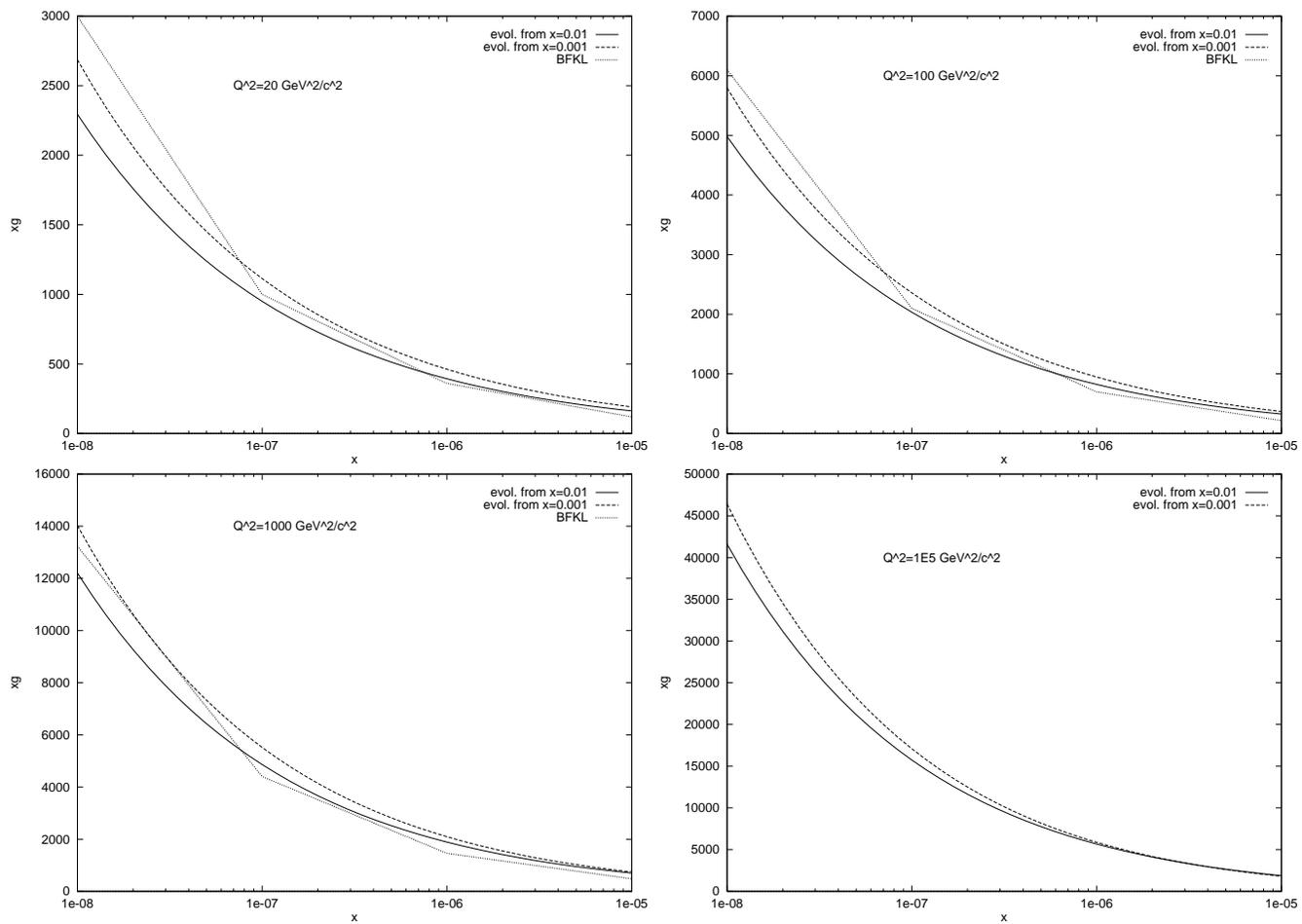
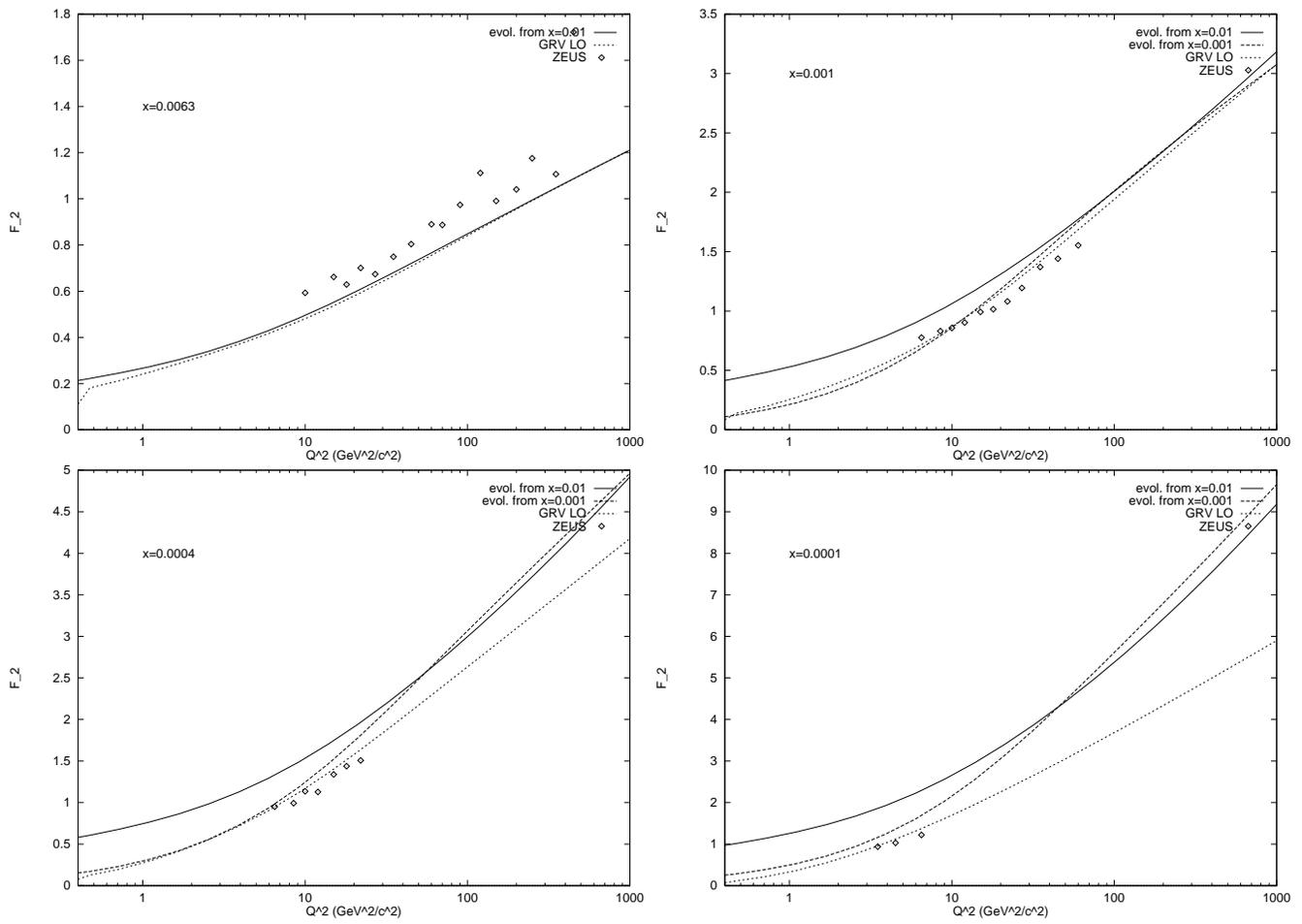
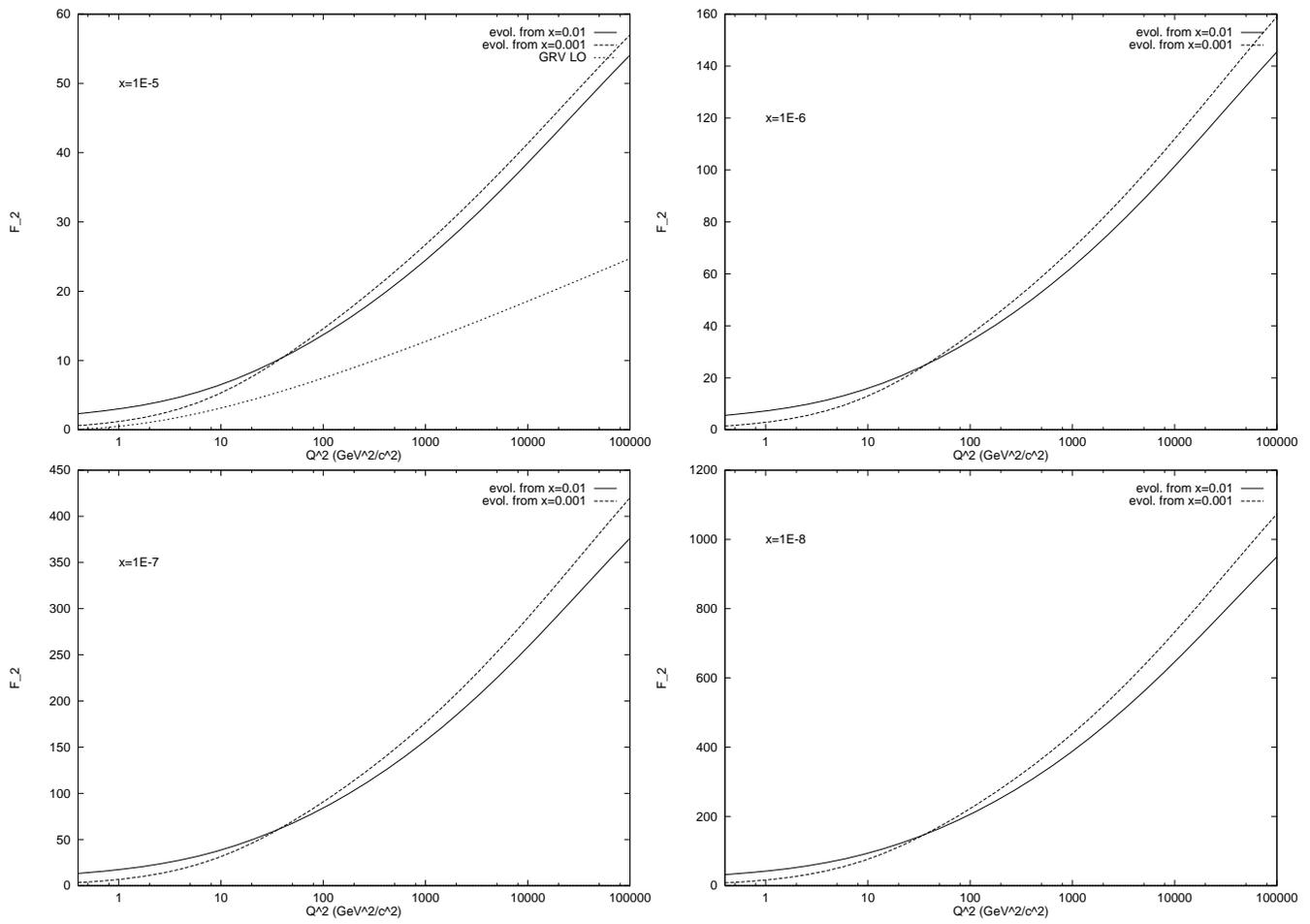
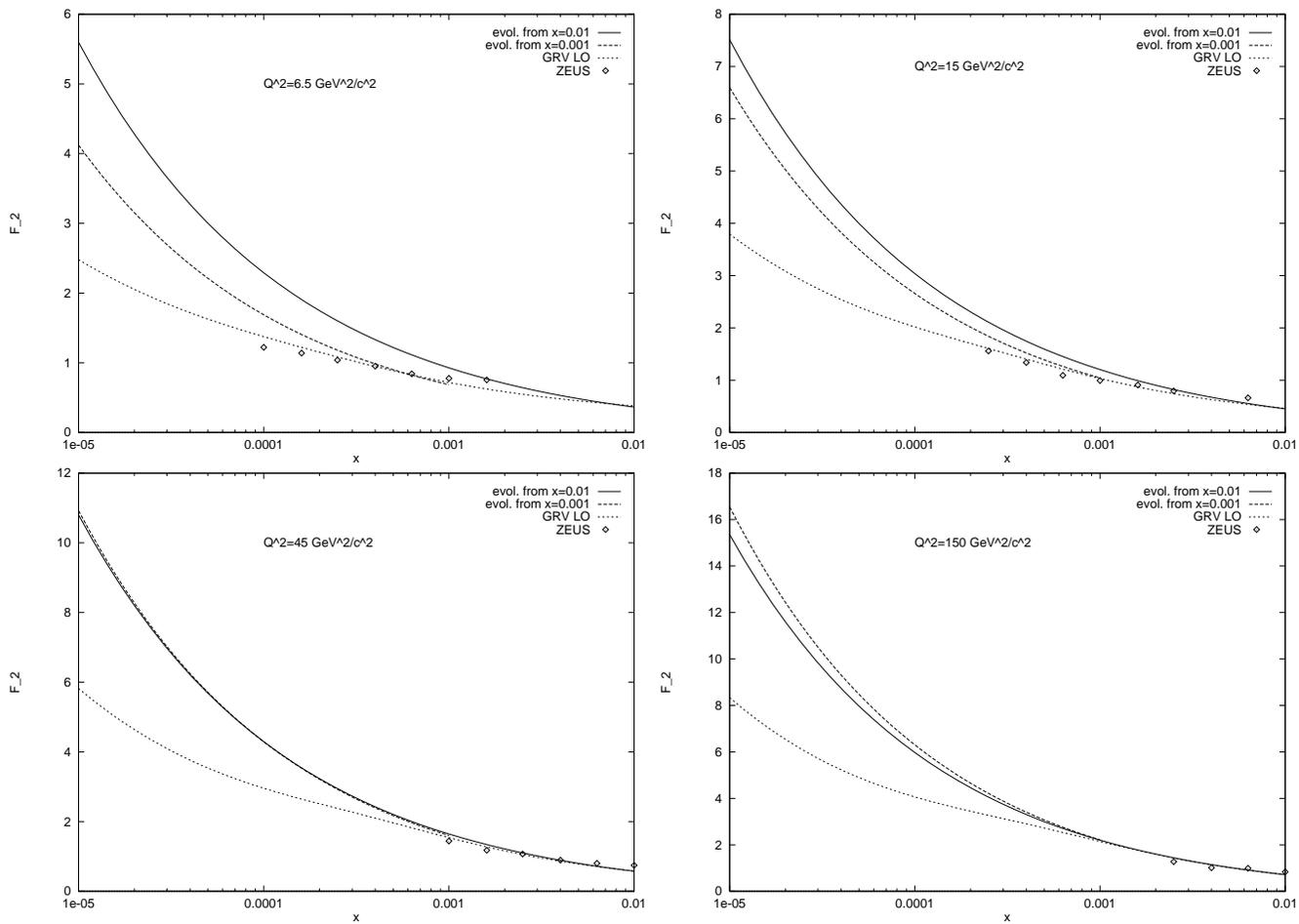
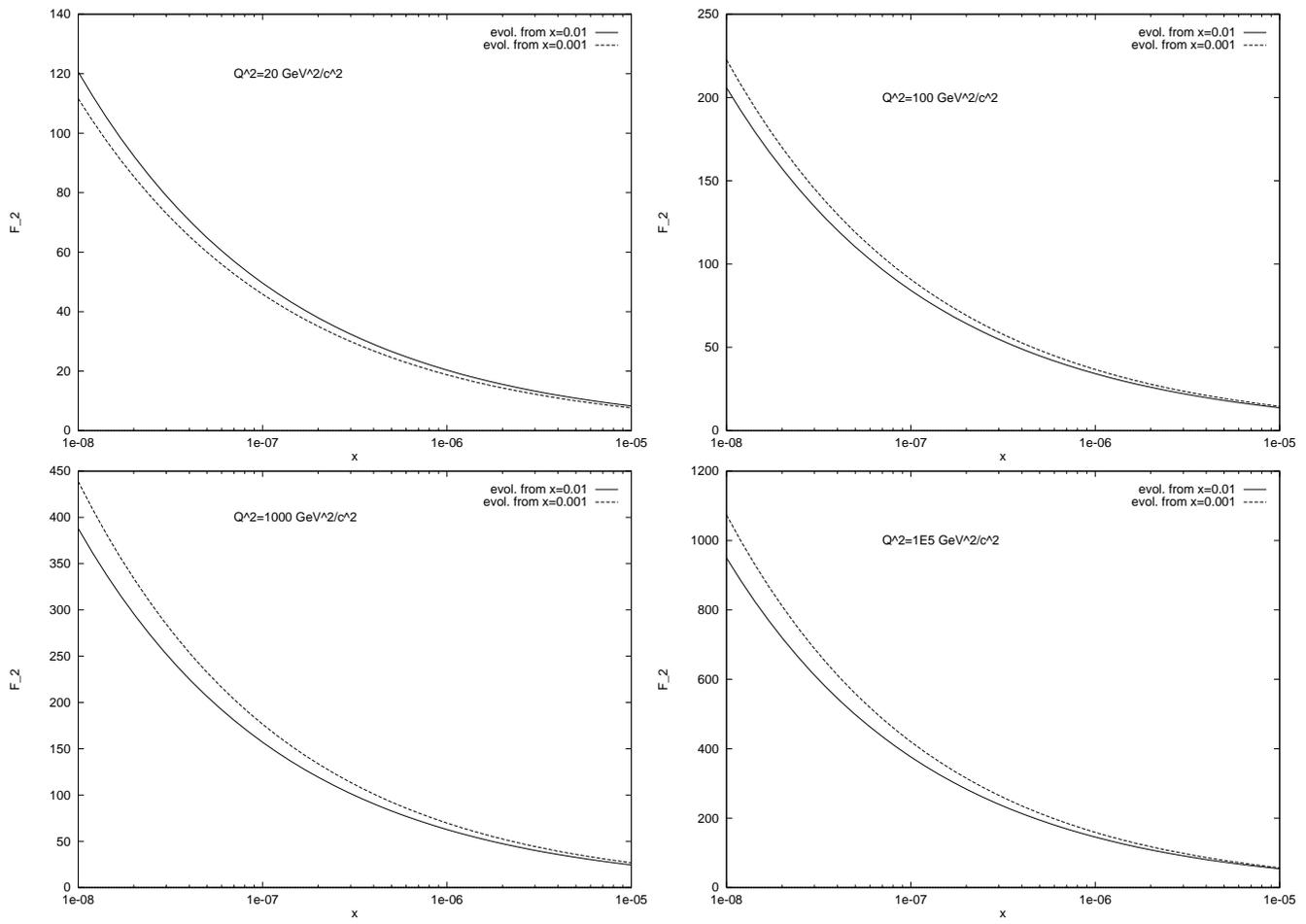
\end{document}